\documentclass[aps,pra,twocolumn,groupedaddress]{revtex4}

\usepackage{graphicx}

\begin{document}

%\preprint{}

\title{Generation of Large
Number-Path Entanglement Using Linear Optics and Feed-Forward}


\author{Hugo Cable}
\email{hcable@lsu.edu}
\author{Jonathan P. Dowling}
\affiliation{ Horace C. Hearne Jr. Institute for Theoretical
Physics, Department of Physics and Astronomy, Louisiana State
University, Baton Rouge LA70803.}

\date{April 4, 2007}

% My shorts:
\def\be{\begin{equation}}
\def\ee{\end{equation}}
\def\bea{\begin{eqnarray}}
\def\eea{\end{eqnarray}}
\newcommand{\ket}[1]{|#1\rangle}
\newcommand{\bra}[1]{\langle#1|}

\begin{abstract}
We show how an idealised measurement procedure can condense
photons from two modes into one, and how, by feeding forward the
results of the measurement, it is possible to generate efficiently
superpositions of components for which only one mode is populated,
commonly called ``N00N states''. For the basic procedure, sources
of number states leak onto a beam splitter, and the output ports
are monitored by photodetectors. We find that detecting a fixed
fraction of the input at one output port suffices to direct the
remainder to the same port with high probability, however large
the initial state. When instead photons are detected at both
ports, Schr\"{o}dinger cat states are produced. We describe a
circuit for making the components of such a state orthogonal, and
another for subsequent conversion to a N00N state. Our approach
scales exponentially better than existing proposals. Important
applications include quantum imaging and metrology.
\end{abstract}

%\pacs{}
%\keywords{}

\maketitle

The fundamental limits to optical detection for metrology and
imaging are quantum mechanical \cite{Giovannetti04}.  Of
particular interest for reaching such quantum limits are
path-entangled states of photons of the form
$\ket{N0}+e^{i\phi}\ket{0N}$, in a basis of photon-number states,
commonly referred to as ``$N00N$'' states. A variety of
applications have been suggested \cite{Kapale05}. For lithography
\cite{Boto00} and microscopy \cite{Teich97}, $N00N$ state light
would be used together with multi-photon absorbers to achieve
enhanced resolution. This is because the de Broglie wavelength for
an $N$-photon state is a factor $1/N$ smaller than the wavelength
associated with the single photon, and the absorption rate scales
linearly with the incident intensity, rather than as the
$N^{\rm{th}}$ power. Regarding applications to precision
metrology, whereby an interferometric setup is used to measure
small phase shifts, $N00N$ states achieve the Heisenberg limit,
for which the phase uncertainty scales as $1/N$
\cite{Bollinger96,Ou97,Boixo06}, and entanglement is a fundamental
requirement for achieving this limit.  It has been rigorously
demonstrated that the cost of improving sensitivity (without using
entanglement) is higher intensities or longer coherence times
\cite{Giovannetti06}. Classically the shot-noise limit applies,
attained for example by laser light, for which the uncertainty
scales as $1/\sqrt{N}$, already a restraint in applications such
as magnetometry \cite{Kominis03} and gyroscopy \cite{Dowling98}.

However, building a source of $N00N$ states beyond two photons is
challenging. Three, four and six photon experiments have been
reported \cite{Mitchell04,Walther04,Resch05}, but only in the
first two references were $N00N$ states generated. In theory a
source could be made using a nonlinear crystal \cite{Sanders89}.
However, the required optical nonlinearity is not readily
available. An alternative is a non-deterministic approach using
linear optics, wherein the desired state is generated on condition
of a specific outcome at photodetectors.  A variety of schemes
have been suggested which typically rely on conditional
destructive interference \cite{Fiurasek02,Zou02,Kok02}. However,
so far none of these scales efficiently, that is they all share
the feature that exponentially decreasing success probabilities
outweigh the possible gains. Noting that quantum algorithms,
exhibiting polynomial and exponential speedups over their
classical counterparts, may be implemented scalably in a
linear-optics approach \cite{KLM01}, we expect that it should be
possible to do better. In this Article we address this challenge
by adapting a conceptually simple measurement procedure.  Our
method is as follows. First, we aim to minimise the negative
effects of back-action in a sequence of detections, whereby
earlier measurements affect the outcomes of later ones. Next, we
engineer output states that closely approximate the ideal case.
Finally, we exploit feed-forward, for which circuits are actively
switched in response to previous photodetections. Feed-forward is
an essential ingredient of linear-optics based quantum computing,
but is not used in previous proposals for engineering $N00N$
states.

We begin by considering the thought experiment depicted in
Fig.~1(a). Here, cavity modes labeled $A$ and $B$ are assumed to
start with a well-defined photon number $N$.  They are each
coupled to an external mode by a weakly transmissive mirror, and
these modes are combined at a 50:50 beam splitter, and then
subject to partial photodetection.   The beam splitter acts to
make the origin of the photons indistinguishable.  When a photon
is registered at the left or right photodetector labeled $D_L$ or
$D_R$, the transformation is given by the Kraus operators
$\hat{L}=(\hat{a}-\hat{b})/\sqrt{2}$ or
$\hat{R}=(\hat{a}+\hat{b})/\sqrt{2}$ respectively (where $\hat{a}$
and $\hat{b}$ are the annihilation operators for modes $A$ and
$B$). To obtain the corresponding probabilities it is necessary to
normalise by the total photon number prior to detection. We
suppose now that a string of detections occur only at $D_r$, say
(by adjusting the path length difference of the cavities between
detections, with a phase shifter, the same state for the cavity
modes can be obtained in every case). After this, the detectors
are removed and the system evolves to a final state with all the
remaining population at the output ports. Denoting by
$\ket{\psi_{AB}}$ the state of modes $A$ and $B$ after $r$ initial
detections, we find that $\left\vert \psi_{AB}\right\rangle
=\hat{R}^{r}\left\vert N\right\rangle \left\vert N\right\rangle
/\sqrt{\left\langle N\right\vert \left\langle N\right\vert \left(
\hat{R}^{\dagger }\right) ^{r}\left(\hat{R}\right)^{r}\left\vert
N\right\rangle \left\vert N\right\rangle }$, normalising to unity.
The probability $P_{\rm cond}$ of finding all the remaining
photons ``condensed'' at the right output port (and none at the
left output port) is as follows,
\begin{eqnarray}
P_{\rm cond}&=&\left\langle \psi_{AB}\right\vert \left(
\hat{R}^{\dagger }\right) ^{S}\left(\hat{R}\right)^{ S}\left\vert
\psi
_{AB}\right\rangle /S! \nonumber \\
&=& \left( ^{2N}C_{N}\right) ^{2}/\left[ 2^{S}\sum_{k=0}^{r}\left(
^{r}C_{k}\right) ^{2}\left( ^{S}C_{N-k}\right) \right], \nonumber
\end{eqnarray}
where $S\equiv2N-r$ denotes the total remaining photon number, $C$
denotes a binomial coefficient, and we assume that $r<N$.
Evaluating the value of $P_{\rm cond}$ numerically for initial
states of increasing size, we find that its value is determined
asymptotically by the proportion of the input that is measured.
For example, setting $r$ either as one quarter or one third of
$2N$ suffices for ${P_{\rm cond}}>0.6$ or ${P_{\rm cond}>0.7}$,
respectively.

\begin{figure}
\fbox{\includegraphics[width=3.75cm]{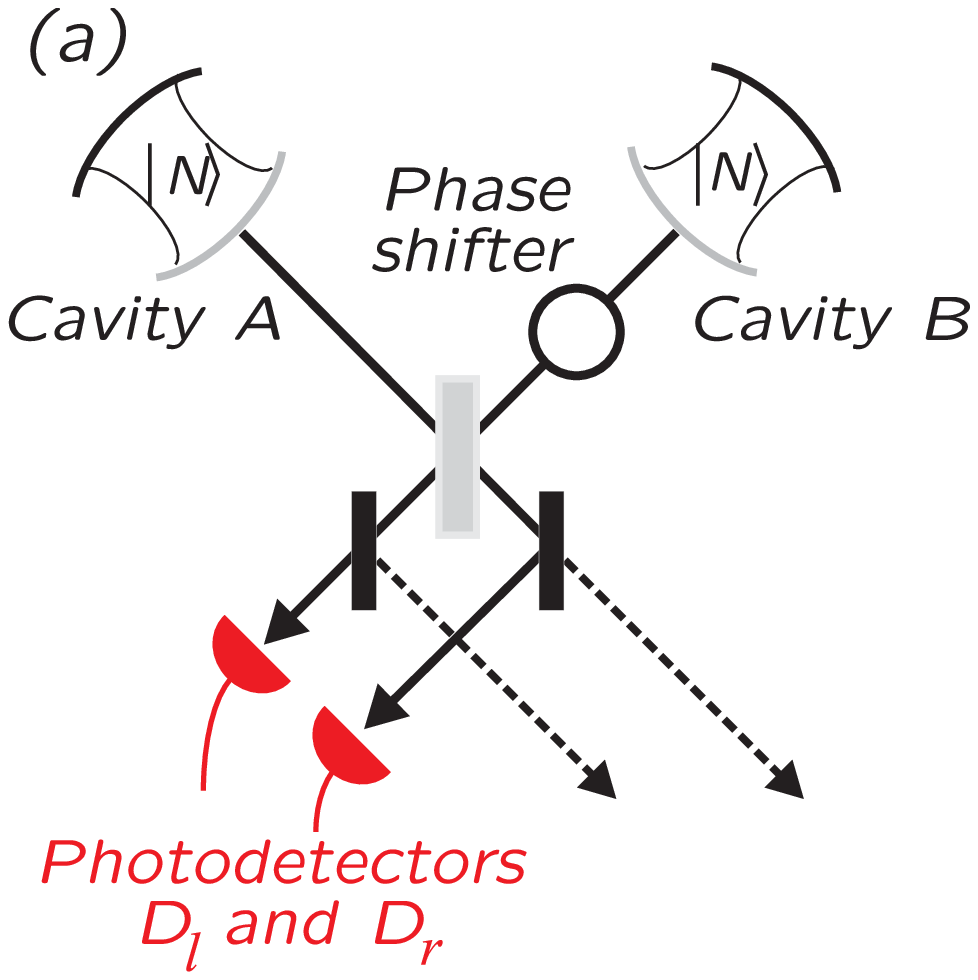}}
\fbox{\includegraphics[width=4.25cm]{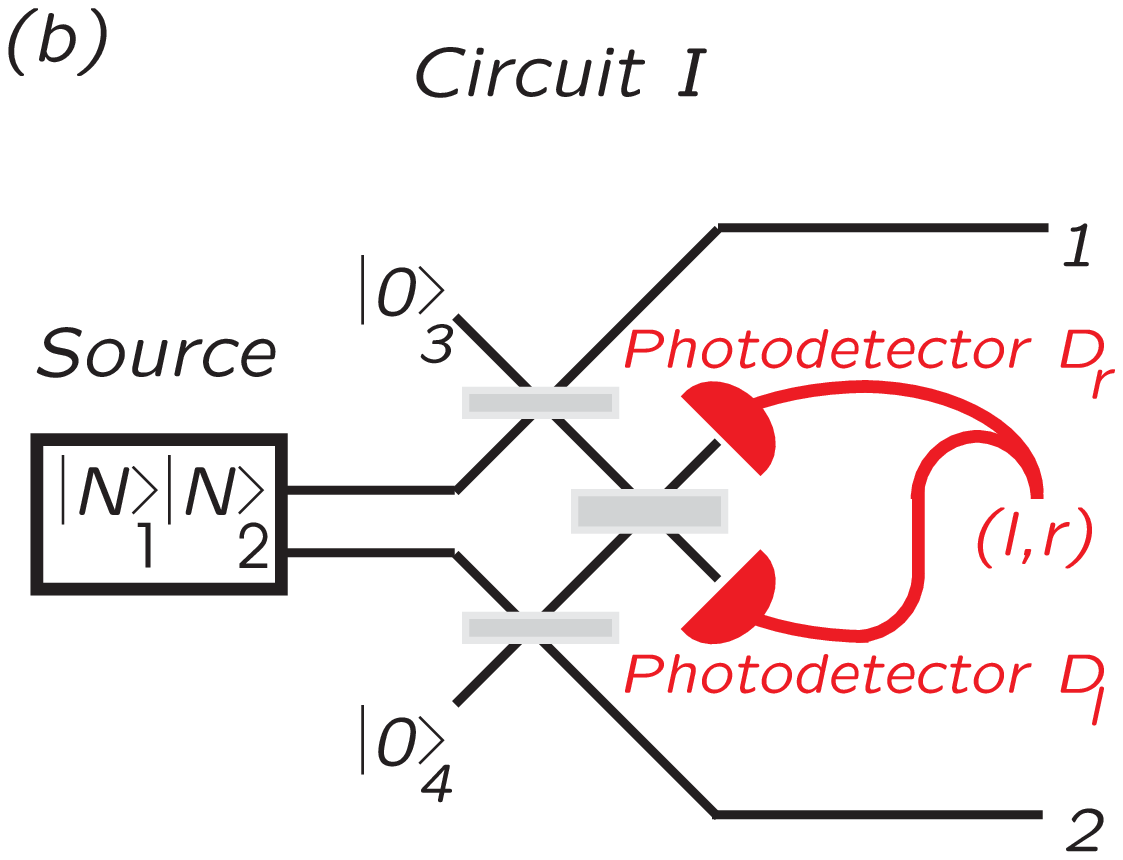}} \caption{Two
measurement-based procedures, inducing relative phase correlations
between the principal modes, each one with $N$ photons at the
start. In (a), population leaks from cavity modes $A$ and $B$ into
external modes, which are combined at a beam splitter. Photons are
detected one at a time by photodetectors $D_l$ and $D_r$. In (b),
all modes are propagating. Beam splitters couple some fraction $f$
from the principal modes one and two into ancillae modes three and
four, which are initially the vacuum. The ancillae are
subsequently combined at a beam splitter, and subjected to
number-resolving photodetection at $D_l$ and $D_r$.}
\end{figure}

We have found for our thought experiment, that later detections
tend to strongly reinforce earlier ones.  Hence, the effect of
measurement back-action here is useful for state engineering, and
in what follows we adapt the measurement process for $N00N$ state
generation.  We divide our analysis into three stages.  First, we
translate our thought experiment into a mathematically equivalent
procedure based purely on linear optics, and consider the general
case for which photons are detected at both photodetectors.  The
localisation phenomena resulting from this measurement process,
 have been studied extensively in the context of the debate over
 the existence of absolute optical coherence in
common quantum-optical experiments \cite{Molmer97}. It has been
demonstrated that well-defined correlations in the relative
optical phase evolve, for the remaining population, and play a
central role in the ongoing dynamics.  Hence, in the second stage
of our analysis, we investigate simple procedures for manipulating
phase correlations, and relate states with well-defined
correlations in the relative phase and $N00N$ states. Finally, we
identify a method based on feed-forward to enable $N00N$ states to
be generated efficiently.

First, we translate our thought experiment into a
mathematically-equivalent procedure, based on linear optics, as
depicted in Fig.~1(b).  We label this optical circuit, Circuit I.
Here all modes are propagating, and a source is assumed to supply
dual Fock states $\ket{N}\ket{N}$ to the principal modes one and
two. Beam splitters of reflectance $f$ couple modes one and two to
ancillae modes three and four, which are combined at a 50:50 beam
splitter. They are then measured by number-resolving
photodetectors labeled $D_l$ and $D_r$, where on average a
fraction $f$ of the input photons are registered. We now consider
the state $\ket{\psi_{l,r}}$ generated in modes one and two after
$l$ photons are registered at $D_l$ and $r$ at $D_r$. Following
Ref.~\cite{Sanders03}, it is convenient to adopt a representation
in terms of coherent states, which are of the form,
$\ket{\alpha}\!\equiv\! \ket{\vert \alpha \vert \exp(i\theta)
}\propto\sum_{k=0}^{\infty}
\sqrt{\vert\alpha\vert^{2k}/k!}\exp\left(ik\theta\right)\ket{k}$
in a basis of Fock states. It has been shown that,
\begin{eqnarray}
\label{eqn:partialloc} \left\vert \psi _{l,r}\right\rangle
\!\!\!&\propto&\!\!\! \int_{0}^{2\pi }\int_{-\pi}^{\pi }d\theta
_{\rm
av}d\Delta\exp \left( -iS\theta_{\rm av}\right) \times \nonumber \\
\!\!\!&&\!\!\! { \Big[} G(\Delta\!-\!\Delta _{0})\!+\!\exp \left(
i\sigma \right) G(\Delta\!+\!\Delta _{0}){ \Big]} \left\vert
\alpha _{1}\right\rangle \! \left\vert \alpha _{2}\right\rangle\!,
\end{eqnarray}
where $S\equiv2N-l-r$, $\alpha_j=\vert \alpha_j \vert
\exp(i\theta_j)$, $\theta_{\rm av}=(\theta_1+\theta_2)/2$ and
$\Delta\equiv\theta_2-\theta_1$. The superposition phase $\sigma$
takes the value $l\pi$, and hence the measurement record must be
known exactly. The scalar function $G(X)$ is given to good
approximation by the Gaussian expression
$\exp\left[-(l+r)X^2/4\right]$. The total photon number is equal
to $S$. There are well-defined correlations in the
relative-optical phase parameter $\Delta$ at values plus and minus
$\Delta_0$, determined only by the ratio of $l$ to $r$. These
correlations are multi-valued whenever photons are registered at
both photodetectors, and a Schr\"{o}dinger cat state is generated.
We can see that cats are generated as a result of the symmetry of
the setup. Specifically, $\hat{L}$ and $\hat{R}$ are invariant
under an exchange of the labeling of the modes, a transformation
which reverses the sign of the relative phase. The generation of
cat states therefore also requires precise phase stability between
the modes. Turning to the source, we see that the state of the
input can be a mixture of the form $\sum_N P_N
\ket{N}\ket{N}\bra{N}\bra{N}$, since $\Delta_0$ is independent of
$N$, and standard linear optical elements obey a superselection
rule for the photon number. Several two-mode squeezing processes
strongly suppress relative number fluctuations, and hence might
serve as practical sources of light described by these mixed
states.

For the second stage in our analysis, we identify the outputs of
Circuit I as examples of quantum reference frames --- reference
frames for a classically defined parameter composed of finite
quantum resources.  Quantum reference frames are subject to
depletion and degradation as they are used, and are currently of
interest for protocols in the field of quantum information, in
which they are regarded as a resource \cite{Bartlett06}.  By
making an analogy to classical phase references we can now
identify simple ways in which states of the form
Eq.~(\ref{eqn:partialloc}) can be manipulated.  For the current
purposes we can assume that a large number of detections have been
performed and define,
\begin{equation}
\label{eqn:psiinfty} \left\vert \psi _{\infty }\left(\Delta
_{0}\right) \right\rangle \propto\int_0^{2\pi}d\theta
\exp\left(-iS\theta\right)\ket{\alpha}\ket{\alpha
\exp(i\Delta_0)},
\end{equation}
where $\alpha=\vert \alpha \vert e^{i\theta}$, for a state with a
total photon number $S$ and a relative phase of $\Delta_0$
(assumed to be normalized). Relative phase correlations between
more than two modes are transitive and are transformed additively
by phase shifters. A phase reference can be extended to additional
modes by combining it with the vacuum at a beam splitter.  As an
example, a 50:50 beam splitter, which we denote here by $U_{\rm
bs}$, beating light in a Fock state with $S$ photons against the
vacuum yields, $U_{\rm
bs}\ket{S}\ket{0}\propto\ket{\psi_\infty(0)}$, and $U_{\rm
bs}\ket{0}\ket{S}\propto\ket{\psi_\infty(\pi)}$. Therefore, we see
that a simple circuit, consisting only of a beam splitter and a
phase shifter, can convert a cat state generated by Circuit I to a
$N00N$ state, whenever the relative phase correlations differ by
$\pi$.  This happens when $l=r$ and $\Delta_0=\pi/2$. We label
this circuit, Circuit III (anticipating an intermediate process
modifying the cat states for the general case).

Before proceeding to the final stage of our analysis, we consider
a simple $N00N$-state generator, that attempts to convert every
cat state generated by Circuit I using Circuit III. This method
might be expected to yield close approximations to $N00N$ states,
whenever the the relative phases of the cat state are close to
plus and minus $\pi/2$. The situation is summarized in Fig.~2(a).
To measure the quality of the output state we adopt the fidelity,
denoting it by $F$. For the measurement-induced condensation,
considered at the start, $F$ takes the same same value as $P_{\rm
cond}$. For schemes generating $N00N$ states, it is necessary to
account for the phase of the superposition, and we define $F=\rm
{max}_{\phi} \Big\vert
\Big(\bra{0S}+\exp\left(-i\phi\right)\bra{S0}\Big) \ket{\psi_{\rm
output}} \Big\vert^2/2$, where $S$ is the total photon number of
the state $\ket{\psi_{\rm output}}$. Evaluating $F$ for our
$N00N$-state generator, when Circuit I generates a cat state with
relative phase components at $\pm \Delta_0$ and total photon
number $S$, we find to first approximation that
$F\sim\cos^{2S}\left[\left(\Delta_0 -\pi/2\right)/2\right]$. As
with other proposals, this scheme in fact scales {\it
exponentially poorly} whenever the relative phase correlations are
less than $\pi$ apart, as is typically the case.  Inspecting the
overlap for different relative phase components, as in
Eq.~(\ref{eqn:psiinfty}) with total photon number $S$, we find
that $\Big\vert\langle
\psi_\infty(\Delta_1)\vert\psi_\infty(\Delta_2)\rangle
\Big\vert=\Big\vert\cos\left[\left(\Delta_2-\Delta_1\right)/2\right]
\Big\vert^S$. The poor scaling can be attributed to the
non-orthogonality of the cat state components.

\begin{figure}
\fbox{\includegraphics[width=3.3cm]{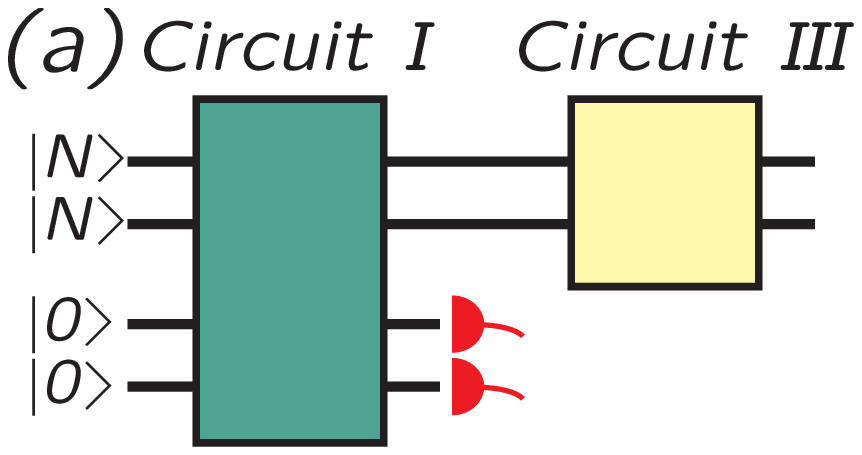}}
\fbox{\includegraphics[width=4.7cm]{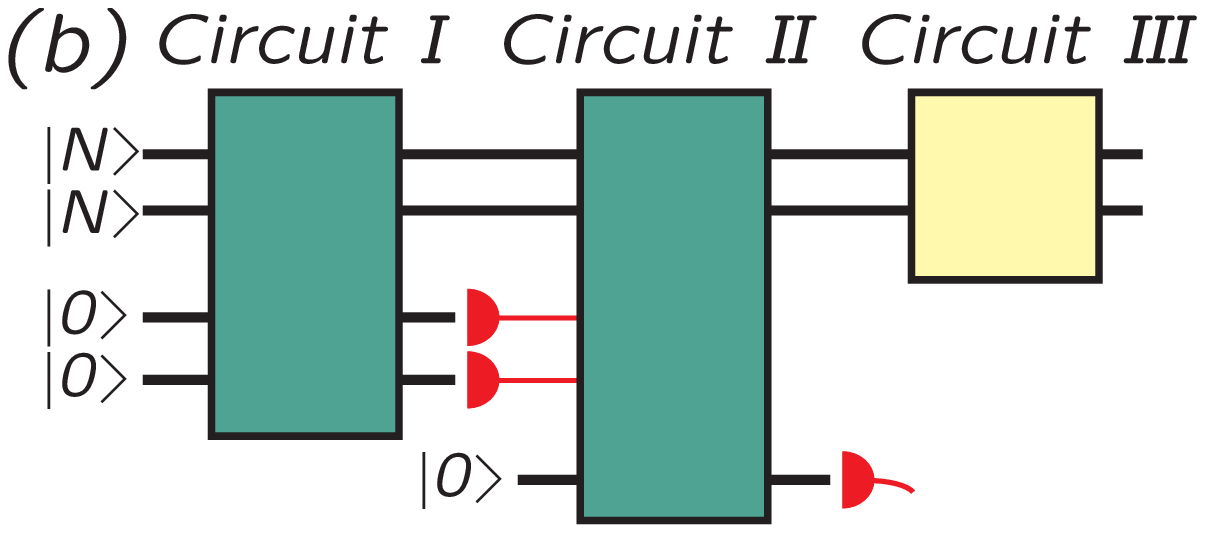}}
 \caption{(a) and (b) illustrate
complete $N00N$ state generators in outline. Circuit I produces
Schr\"{o}dinger cat states with two relative phase components
non-deterministically (green). The generators are terminated by a
fixed unitary (yellow) circuit consisting of beam splitters and
phase shifters.  In (b) the cat states are corrected, using an
additional measurement process conditioned on the previous
detection outcomes.}
\end{figure}

We now proceed to the final stage of our analysis.  Our previous
$N00N$-state generator is effective when Circuit I generates cat
states with components which are orthogonal. However, this occurs
with low probability.  It is not possible to improve the situation
with any combination of (idealised) beam splitters and phase
shifters, since these implement unitary transformations. Hence, we
now devise a circuit, labeled Circuit II, to make input cat
components orthogonal, using additional processes of measurement
and feed-forward.  This more sophisticated scenario is depicted in
Fig.~2(b). To identify a suitable circuit, we investigate how a
beam splitter transforms phase references, starting with two
classical fields. Here the field in each mode is represented by a
complex number, with the square amplitude corresponding to the
intensity, and the phase to the optical one. A 50:50 beam
splitter, configured so as not to impart additional phase shifts
to the modes, outputs two classical fields described by the sum
and difference of the values for the inputs (altering both the
square amplitudes and the phases).  If the input has a relative
phase of $0$ or $\pi$, and equal intensities for each mode, the
population is transferred entirely into one mode. On the other
hand, if the input has a relative phase of plus or minus $\pi/2$,
and equal intensities in each mode, the relative phase and
intensities are preserved. Moving to the quantum case, we consider
the action of the beam splitter for a state, defined as in
Eq.~(\ref{eqn:psiinfty}), with a relative phase of $\Delta_0$, a
total photon number $S$, and an intensity $S/2$ in each mode.
Computing the final state explicitly, we find a scenario similar
to the classical case,
\begin{eqnarray}
\label{eqn:bsonquantumphasereference} U_{\rm
bs}\ket{\psi_\infty}\!\!&\propto&\!\!\int^{2\pi}_0 d\theta
\exp\left(-iS\theta\right) \times \nonumber \\
\!\! && \!\!\ket{\sqrt{I_1} \exp\left(i\theta\right)} \ket{
\sqrt{I_2} \exp\left[i\left(\theta\pm \pi/2\right)\right]}\,.
\end{eqnarray}
This has an intensity $S I_1/\left(I_1+I_2\right)=
\left[1-\cos(\Delta_0)\right]/2$ in mode one and $S
I_2/\left(I_1+I_2\right)=S \left[1+\cos(\Delta_0)\right]/2$ in
mode two, and a relative phase of plus $\pi/2$ when
$0<\Delta_0\leq\pi/2$ and of minus $\pi/2$ when
$-\pi/2\leq\Delta_0<0$ (we consider cases for which the intensity
is increased in favour of mode two).

The symmetry of the beam splitter transformation makes it useful
for altering the cats generated by Circuit I, so that the relative
phases are different by $\pi$.  However, it creates a difference
in the intensities between the modes. To correct this, we propose
beating mode two against the vacuum, so as to to move the
difference of the intensities to an ancillary mode, which can be
removed by a photodetection.  This method depends critically on
feeding forward the result of the detections performed by Circuit
I, so that a variable beam splitter can be set according to the
value of $\Delta_0$.  A variable beam splitter can be implemented
with 50:50 beam splitters and variable phase shifters. The cost of
correction is a decrease in the total photon number, which varies
non-deterministically. As can be seen from
Eq.~(\ref{eqn:bsonquantumphasereference}), a fraction of
$\cos\left(\Delta_0\right)$ of the photons are lost on average.
Overall, Circuits I through III constitute a complete $N00N$-state
generator.  Additional mathematical analysis is given in the
supplementary online text. Runs for which Circuit I fails to
generate a Schr\"{o}dinger cat state, or too many photons are lost
in the detection process are discarded.  The fidelities at the
output are, on average, $0.87$, $0.94$ or $0.98$, when a fraction
of one third, one half, or two third respectively of the input
photons are detected by Circuit I. Higher fidelities are possible
when the photon number at the input is small.  If allowance is
made for sufficient input photons to be detected by Circuit I, and
a further half to be detected in Circuit II, the probability of
failure is not too large.

Finally, we suggest some possibilities for experimental
implementation. For the source, we propose an optical parametric
oscillator setup for which the two mode squeezed output of an
optical parametric amplifier is enhanced by a cavity
\cite{Zhang02}. Note, however, that the current purposes require
twin beams of a much lower intensity than is typical in many
experiments, and that the beams must be rendered frequency
degenerate. Techniques of feed-forward and photodetection are
being developed with a view to quantum information technologies
\cite{Kok07,Prevedel07}. For the source, an important problem is
imperfect correlation between the modes. If, for example, two
independent lasers of equal intensity provide the input, the
scheme generates the intended relative phase correlations, but no
entanglement \cite{Cable05}. Photodetectors are subject to loss
and dark counts.  Losses will act to degrade the source, reducing
the relative number correlation and increasing the uncertainty in
the total photon number.  Dark counts are more problematic, mixing
over the phase for the superposition in
Eq.~(\ref{eqn:partialloc}). An alternative suggestion is using
trapped bosonic atoms. One possibility might be to work in a
regime for which the atomic wave-packets are much longer than the
typical scattering length, as proposed in Ref.~\cite{Popescu06}.
Another is to use Bose-Einstein condensates, for which a variety
of coherent operations have been demonstrated. Number-resolved
condensates might be obtained from the Mott Insulator phase, while
relative-number squeezing can be achieved by different techniques.

In conclusion, we have proposed for the first time a linear-optics
based scheme that generates large $N00N$ states efficiently, the
photon number at the output scaling with that of the source
--- all the while maintaining high fidelities, high success
probabilities and a fixed number of circuit components. As well as
being of immediate interest for a range of applications, our
results have connections with other topics. For example, the
scaling we derive for our measurement-induced condensation
procedure is of relevance to the study of the interference of
light from independent sources and localizing relative optical
phase, phenomena with analogs in different physical systems
\cite{Rau03}. We have left as an open question the extent to which
the scaling can be attributed to Bose statistics.  Regarding our
$N00N$-state generators, the creation of macroscopic entangled
states is of interest for exploring the quantum-classical
transition. Finally, our study of Schr\"{o}dinger cat states may
have application to quantum computing, where Schr\"{o}dinger cat
states, defined for one mode only, have been proposed to encode
qubits, which may be manipulated using standard experimental
techniques \cite{Jeong05}.

\vspace{1cm}
\begin{center}
{\bf Acknowledgements}
\end{center}
The authors would like to acknowledge support from the Hearne
Institute, the Army Research Office, and the Disruptive
Technologies Office.  H. C. would like to thank Terry Rudolph,
Ryan Glasser, Sonja Daffer and Yuan Liang Lim for helpful
discussions.

\clearpage

\begin{center}
{\bf Supplementary Material: Methods}
\end{center}

In these supplementary notes, we provide further analysis of our
$N00N$-state generator, consisting of Circuits I, II and III, as
depicted in outline in Fig.~2(b). First, we specify notation for
beam splitters, phase shifters and states with well-defined
relative phase correlations. For the lossless beam splitter, we
choose a notation which makes explicit the ``rotation'' performed
by such a device. A beam splitter with transmittance $\tau$ and
reflectance $\left(1-\tau\right)$ acts to transform the
annihilation operators $\hat{o}_j$ for modes labeled $j$,
according to the relations,
\[
\left(%
\begin{array}{c}
  \hat{o}_1 \\
  \hat{o}_2 \\
\end{array}%
\right) \longrightarrow
\left(%
\begin{array}{cc}
  \cos(\gamma) & -\sin(\gamma) \\
  \sin(\gamma) & \cos(\gamma) \\
\end{array}%
\right)
\left(%
\begin{array}{c}
  \hat{o}_1 \\
  \hat{o}_2 \\
\end{array}%
\right)
\]
with angular parameter $\gamma$, where
$\tau=\cos^2\left(\gamma\right)$ and $0\leq\gamma\leq\pi/2$.  We
denote this transformation $U_{\rm bs}\left(\gamma \right)$, and
we include, where necessary, phase shifts of $\chi$ at the input
port and $-\chi$ at the output port of the first mode, so that
$U_{\rm
bs}\left(\gamma,\chi\right)\equiv\exp\left(\gamma\exp(i\chi)\hat{o}_1
\hat{o}_2^\dagger-\gamma\exp(-i\chi)\hat{o}_1^\dagger
\hat{o}_2\right)$. For example, $U_{\rm
bs}\left(\arccos{\left(\sqrt{\tau}\right)},\pi/2\right)$
corresponds to a symmetric beam splitter.  We denote a phase shift
transformation on mode $j$, $\exp\left(i \hat{o}_j{}^\dagger
\hat{o}_j \chi\right)$, by $U_{\rm ps}\left(\chi\right)$. For a
state defined, as in Eq.~(2), with a total photon number $S$ and
relative phase $\Delta_0$, it is helpful to incorporate a phase
factor $\exp\left(-iS\Delta_0/2\right)$ into the normalisation
(making the definition symmetric between the modes). We then adopt
the following notation for a normalised Schr\"{o}dinger cat state,
\[
\left\vert \psi_{\rm cat}\left(\Delta_{0},\Lambda \right)
\right\rangle \propto \left\vert \psi _{\infty }\left( \Delta
_{0}\right) \right\rangle +\exp\left(i\Lambda\right)\left\vert
\psi _{\infty}\left(-\Delta_0\right)\right\rangle,
\]
having components with relative phases plus and minus $\Delta_0$,
a phase for the superposition $\Lambda$ (with the overall
normalisation constant assumed positive).

Next, we elaborate on the sequence of operations performed by our
$N00N$-state generator.  We assume the final state should have at
least $P$ photons, and that the correlations in the relative phase
are ideal. For the first step, Circuit I, depicted in Fig.~1(b),
implements the transformation,
\[
\left\vert l,r\right\rangle \!\! \left\langle l,r\right\vert
_{3,4}U_{\rm bs}\left( \frac{\pi}{4}\right) _{3,4}\!\!U_{\rm
bs}\!\!\left[ {\rm asin} \left( \!\sqrt{f}\right) \right]
_{1,3}\!\!U_{\rm bs}\!\!\left[ {\rm asin} \left( \!\sqrt{f}\right)
\right]_{2,4}\!,
\]
conditioned on the detection of $l$ photons in mode $3$ and $r$
photons in mode $4$. A dual Fock state from the source evolves to
a cat state according to,
\[
{\rm Circuit\, I:} \ket{N}_1\ket{N}_2\ket{0}_3\ket{0}_4
\longrightarrow \left\vert \psi_{\rm cat}\left( \Delta_{0},l\pi
\right) \right\rangle_{1,2}.
\]
This cat state has relative phase components with values plus and
minus $\Delta _{0}\equiv 2\arccos \left[ \sqrt{r/(l+r)}\right]$,
and total photon number $2N-l-r$. Runs for which $l$ or $r$ are
zero must be discarded.  It has been shown that values for the
relative phase are generated with approximately equal frequency
across the range \cite{Cable05}, and hence these failure events do
not affect the scaling of the generator.

Next, Circuit II acts to transform the relative phase
correlations, to plus and minus $\pi/2$, in every case.  When
$r\geq l$, the relative phase correlations lie in the range
$[-\pi/2,\pi/2]$, and a 50:50 beam splitter acting on the
principal modes corrects the relative phase correlations, while
increasing the intensity in mode two (and decreasing it in mode
one). To achieve the same outcome when $l<r$, we suppose that a
phase shift of $\pi$ is applied in advance (on either mode).  This
transforms the cat state generated by Circuit I as,
\[U_{\rm
ps}\left( \pi \right) \left\vert \psi _{\rm cat}\left( \Delta
_{0}(l,r),l\pi \right) \right\rangle \propto \left\vert \psi _{\rm
cat}\left( \Delta _{0}(r,l),r\pi \right) \right\rangle.
\]
Next, a beam splitter, with transmittance $\left[ 1-\cos \left(
\Delta _{0}\right) \right] /\left[ 1+\cos \left( \Delta
_{0}\right) \right]$, transfers the difference of the intensities
to the ancillary mode five.  A circuit for implementing the
variable beam splitter is given by the relation,
\begin{eqnarray}
&& U_{\rm bs}\left( \gamma \right) _{2,5}\equiv  \nonumber \\
&& U_{\rm bs}\left( \pi /4,\pi /2\right) _{2,5}U_{\rm ps}\left(
\gamma \right) _{5}U_{\rm ps}\left( -\gamma \right) _{2}U_{\rm
bs}\left( \pi /4,-\pi /2\right) _{2,5}. \nonumber
\end{eqnarray}
A photodetector measures $Q$ photons in mode $5$.  Overall,
Circuit II implements the transformation,
\[\left\vert Q\right\rangle
\left\langle Q\right\vert _{5}U_{\rm bs}\left\{ \arccos \left[
\tan \left( \Delta _{0}/2\right) \right] \right\} _{2,5}U_{\rm
bs}\left( \pi /4\right) _{1,2}.
\]
The cat state evolves as,
\begin{eqnarray}
\! && {\rm Circuit\, II:} \nonumber \\
\! && \left\vert \psi_{\rm cat}\left( \Delta _{0},l\pi \right)
\right\rangle \! \longrightarrow \! \left\vert \psi _{\rm cat}
\left[ \pi /2,l\pi+(2N-l-r-Q)\pi /2 \right] \right\rangle.
\nonumber
\end{eqnarray}
We derived a full probability distribution for the outcomes of the
photodetection performed by Circuit II,
\begin{widetext}
\begin{eqnarray}
{\rm Prob}(Q=0,\cdots,S-1)\!&=&\!\frac{1}{ 1+\left( -1\right)
^{l}\cos ^{S} \left( \Delta _{0}\right)} \,\,{^{S}C_{Q}}\,
 \left[ 1\!-\!\cos \left( \Delta _{0}\right) \right]
^{S-Q}\,\cos ^{Q}\left( \Delta _{0}\right) \nonumber \\
{\rm Prob}(Q=S)\!&=&\!\frac{1}{1+\left( -1\right) ^{l}\cos
^{S}\left(\Delta_0\right)} \frac{\left[ 1+\left(
-1\right)^{l}\right] ^{2}}{2} \cos ^{S}\left(\Delta_0\right)
\,\,,\nonumber
\end{eqnarray}
\end{widetext}
where $S=2N-l-r$ is the total photon number prior to detection,
and $C$ denotes a binomial coefficient. This probability
distribution is approximately binomial, and the expected number of
detections is $S\cos\left(\Delta_0\right)$.  If too many photons
are lost in Circuits I and II the run must be aborted.  The
probability of this can be made small by taking $P \simeq N (1-f)
$.  In principle, excess photons can be removed by an additional
process, similar to Circuit I.

Finally, Circuit III implements the unitary transformation,
\[
U_{\rm bs}\left( \pi /4,\pi \right) _{1,2}U_{\rm ps}\left( \pi
/2\right) _{2}.
\]
The corrected cat state evolves as,
\begin{eqnarray}
&&\!\!\!\!\!\!\! {\rm Circuit \, III:} \nonumber \\
&&\!\!\!\!\!\!\! \left\vert \psi _{\rm cat}\left( \pi /2,l\pi
\!+\!(2N\!\!-\!\!l\!\!-\!\!r\!\!-\!\!Q)\pi /2\right) \right\rangle
\!\longrightarrow\! \left\vert
P,0\right\rangle\!+\!(-1)^{l}\left\vert 0,P\right\rangle,
\nonumber
\end{eqnarray}
 yielding the desired $N00N$ state, with
$P=2N\!-\!l\!-\!r\!-\!Q$. It may be noted that the superposition
phase for the $N00N$-state at the output depends on the
measurement record at the photodetectors. When $l<r$, the
additional phase shift in Circuit II causes this phase to be
$r\pi$ rather than $l\pi$.

Next, we estimate the fidelities of the states produced by our
$N00N$ state generator, and clarify its behaviour for large photon
number.  To do this, we first compute the fidelity for one
component of a cat state generated by Circuit I, which we denote
by $\ket{\psi_G(\Delta_0)}$. We assume, as in Eq.~(1), that the
function $G(X)$, describing the localisation of the relative
phase, assumes its Gaussian asymptotic form.  Note that the rate
of localisation is faster when phase correlations evolve at more
than one value.  We assume that the state at the input is the dual
Fock state $\ket{N}_1\ket{N}_2$, and that a total of $D=l+r$
detections have occurred. Then,
\begin{eqnarray}
 F&\sim&\Big \vert \langle \psi_{\rm
G}(\Delta_0) \vert
\psi_\infty\left(\Delta_0\right) \rangle \Big\vert^2 \nonumber \\
 &\simeq&\frac{\left[ \int^{\pi/2}_{-\pi/2} d\Delta \cos^{S}
 \left(\frac{\Delta}{2}\right)
 \exp \left(
-\frac{D\Delta^2}{4} \right) \right]
^{2}}{\int^{\pi/2}_{-\pi/2}\int^{\pi/2}_{-\pi/2}d\Delta
d\Delta^{\prime}\cos ^{S}\left( \frac{\Delta -\Delta ^{\prime
}}{2}\right) \exp \left[ -\frac{D}{4} \left( {\Delta^{\prime}}
^{2}+\Delta
 ^{2}\right)\right]  } \nonumber \\
&\rightarrow&\sqrt{1-\frac{1}{\left[ 2\left( \frac{2N}{S}\right)
-1\right] ^{2}}}\,\,\,, \nonumber
\end{eqnarray}
 where $S=2N-D$ is the total photon number.  This result was
derived assuming that $S$ and $D$ are not small.  The value for
the fidelity depends only on the ratio of detections to input
photons. For example, when $D/2N$ is one half, $F=0.94$, and when
$D/2N$ is two thirds, $F=0.98$.  To verify this result, we
computed numerically exact values for the fidelity, $\Big\vert
\langle \psi_{\rm l,r} \vert \psi_{\rm
cat}\left(\Delta_0(l,r),l\pi\right) \rangle \Big\vert^2$, for a
range of states $\ket{\psi_{l,r}}$ generated by Circuit I. For
input state $\ket{3}_1\ket{3}_2$, the fidelity is $0.94$ for
$(l,r)=(1,2)$ and $(2,1)$, and anomalously it is $1$ for
$(l,r)=(1,1)$.  For input state $\ket{5}_1\ket{5}_2$ the values
are $0.94$ and $0.96$ when $l+r=5$, and range from $0.96$ to $1$
when $l+r=6$, while for input state $\ket{15}_1\ket{15}_2$ the
values range from $0.92$ to $0.96$ when $l+r=15$, and from $0.96$
to $0.99$ when $l+r=20$.

Finally, we performed a complete numerical simulation of the
$N00N$-state generator, to verify that Circuits I, II and III work
together as predicted.  In particular, it was necessary to check
that Circuits II and Circuit III function as expected when the
relative phase correlations for the cat states are not perfectly
well-defined. The results are shown in Fig.~3. Each point in the
plots corresponds to a particular choice of input state, and
measurement by Circuit I. The height corresponds to the expected
photon number for the output $N00N$ state, and the color to its
fidelity. Averages are taken over all possible outcomes to the
third photodetection performed by Circuit II.  For comparison, the
mesh shows the predictions of the preceding analysis.  Good
agreement is seen between these analytical predictions and the
numerical results. However, inspection of individual outcomes in
Circuit II reveals that the high fidelities are not maintained in
every case. Roughly speaking, improbable outcomes were often found
to have low fidelity.

\newpage

\begin{figure}[t]
\includegraphics[height=8cm,width=12cm]{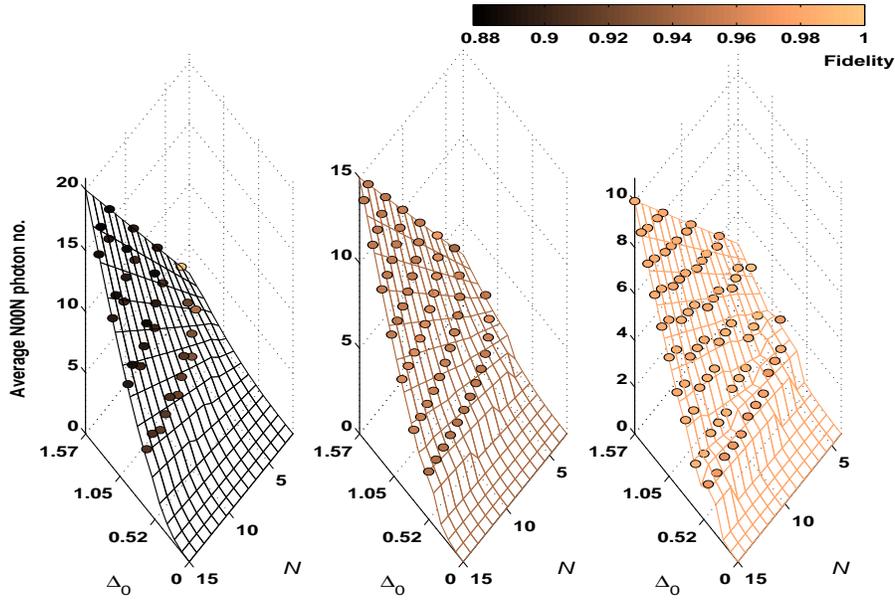}
\caption{Fidelities (color) and photon number (vertical axis) are
displayed for outputs of our $N00N$ states generator. Each point
corresponds to a possible outcome to Circuit I, for which $D$
photons are detected. Input states $\ket{N}_\ket{N}_2$ are
considered for $N$ up to $15$. Going from left to right, $D/2N$ is
one third, one half and two thirds.}
\end{figure}
\end{document}